\documentclass[usenatbib,times]{mn2e}
\usepackage{graphicx}
\usepackage{epsf}

\def\gbm{{{\it Fermi}/GBM}}

\def\H0{{\rm ~km~s^{-1}~Mpc^{-1}}}

\begin{document}

\title[The width of GRB spectra]{The width of gamma-ray burst spectra}

\author[Axelsson \& Borgonovo]{Magnus Axelsson,$^{1,2}$\thanks{email: magnusa@astro.su.se}
and Luis Borgonovo$^{2}$\\
$^{1}$Oskar Klein Center for CosmoParticle Physics, Department of Physics, Stockholm University, SE-106 91 Stockholm, Sweden\\
$^{2}$Department of Astronomy, Stockholm University, SE-106 91 Stockholm, Sweden\\
}

\date{Accepted 2014 December 13.  Received 2014 December 12; in original form 2014 November 14}

\pagerange{\pageref{firstpage}--\pageref{lastpage}} \pubyear{2002}

\maketitle

\begin{abstract}
The emission processes active in the highly relativistic jets of gamma-ray bursts (GRBs) remain unknown. In this paper we propose 
a new measure to describe spectra: the width of the $EF_E$ spectrum, a quantity dependent only on finding a good fit to the data. 
We apply this to the full sample
of GRBs observed by {\gbm} and {\it CGRO}/BATSE. The results from the two instruments are fully consistent. We find that the median 
widths of spectra from long and short GRBs are significantly different (chance probability $<10^{-6}$). The width does not correlate with 
either duration or hardness, and this is thus a new, independent distinction between the two classes. Comparing the measured spectra 
with widths of spectra from fundamental emission processes -- synchrotron and blackbody radiation -- the results indicate that 
a large fraction of GRB spectra are too narrow to be explained by synchrotron radiation from a distribution of electron energies:
for example, 78\% of long GRBs and 85\% of short GRBs are incompatible with the minimum width of standard slow cooling synchrotron 
emission from a Maxwellian distribution of electrons, with fast cooling spectra predicting even wider spectra. Photospheric emission can 
explain the spectra if mechanisms are invoked to give a spectrum much broader than a blackbody.

\end{abstract}

\section{Introduction}

Although studied for several decades, the emission mechanisms in gamma-ray burst (GRB) prompt emission still remain the subject of debate.
It is by now agreed that GRBs are connected to a relativistically expanding outflow or jet \citep{woo93}, but the mechanisms for converting kinetic energy 
to observable radiation remain unknown. A number of different models have been proposed, and the radiative processes involved vary from synchrotron 
\citep[e.g.,][]{rm92}, Compton scattering \citep{bel10} to thermal emission from a photosphere \citep{mes02}. 

The key component in deciphering the physics behind GRBs are their spectra. Despite the fact that lightcurve structure and duration vary 
greatly between events, the overall spectral shape is remarkably similar. This has led to the use of the empirical Band function \citep{band}
to fit the spectra. The function is a smooth joining of two power-laws (with low- and high-energy index $\alpha$ and $\beta$, respectively), and 
provides a good fit to most GRB spectra. However, it lacks any physical motivation and the derived parameters have no immediate interpretation.

The success of the Band function lies in the fact that it can adequately fit the shape of most GRB spectra. The function can mimic both thermal 
and non-thermal processes, and various attempts have been made to connect its parameters to physically derived properties in the models. Most 
such analysis has been done at the level of individual GRBs; however, some attempts have been made to look at the distribution
of parameters, most notable when comparing the low-energy $\alpha$ index to that of the synchrotron spectrum, leading to the so-called ``line 
of death problem" \citep{preece}.

It was quickly found that GRBs could be divided into two classes, long/soft and short/hard, based on their duration and hardness \citep{kou93}. The 
two classes are believed to be the result of different progenitors: long GRBs from collapsing massive stars and short GRBs from the merger of two 
neutron stars or a neutron star and a black hole. Indeed, supernova signatures have been found both in the afterglow spectrum and lightcurve of 
long GRBs \citep[see, e.g.,][and references therein.]{hb12}. Given these two very different progenitors it is remarkable how similar the prompt emission 
spectra are, with short GRBs on 
average having slightly harder spectra than long ones. Any additional property in which the two classes differ would therefore provide welcome 
information to help probe the differences in progenitor mechanisms.

In this paper we present results based on the two largest samples of broad-band GRB spectra: the Burst and Transient Source Experiment (BATSE) on the 
{\it Compton Gamma-ray Observatory} ({\it CGRO}) and the the Gamma-ray Burst Monitor \citep[GBM;][]{gbmpaper} onboard the {\it Fermi Gamma-ray 
Space Telescope}. We will use the standard peak flux spectral fits to constrain the emission mechanisms based on the overall {\it shape} of the spectra. This 
can be derived from the Band function fits irrespective of their lack of physical motivation, and thus provide a model independent test. We will focus on 
the width of the $EF_E$ spectra, as this is a well-defined and easily calculated property. We begin by describing the data analysis and definition of width 
in Sect.~2. Thereafter we present our results, as well as comparisons to spectra from basic radiative processes, in Sect.~3. Finally, we discuss our results.

\section{Data analysis}

In this study we use the peak flux spectral fits presented in the 2nd GBM catalog \citep{gru14}. The catalog contains 943 GRB spectra, detected by {\gbm} in 
the energy range of 8\,keV to 40\,MeV between 2008 and 2012. As part of the standard procedure, all spectra are fit using the Band function and we use these 
parameters to determine the shape of the spectrum. Although the Band function does not provide any clues as to the physical processes behind the emission, 
in this step we are focused on finding a good description of the spectral shape in the energy range where most of the power is radiated. We choose the spectral 
fits made of the peak flux spectra in order to minimize the effect of spectral evolution. The integration time for these spectra are 1.024\,s for long GRBs and 
64\,ms for short GRBs \citep{gru14}.

Augmenting the {\gbm} results, we also analyse 1970 GRB observations from BATSE. Not only does this make our results more instrument independent, but 
the energy ranges and the mean energy of the spectra are slightly different (the BATSE range is 20--600\,keV). We use the sample presented in \citet{gol13}. 
As in the case of {\gbm}, we analyse peak flux spectra (the integration time is 2.048\,s for all BATSE spectra) and the methodology used is the same for both 
catalogs.

In our analysis, we separate long and short GRBs. This is done based on the $T_{90}$ duration, i.e. the time during which 90\% of the emission is measured.
Following standard classification we separate long and short GRBs at 2\,s.

\subsection{Definition of width}

As measurement of the width of the spectra, we use the full width half maximum (FWHM) of the $EF_E$ versus $E$ spectra. As the absolute width
is dependent on the location of the spectral peak, we define the width $W$ as
\begin{equation}
W=\log{\left(\frac{E_2}{E_1}\right)}\, ,
\label{widthdef}
\end{equation}

\noindent where $E_1$ and $E_2$ are the lower and upper energy bounds of the FWHM range, respectively. With this measure, $W$ only depends 
on the Band function parameters $\alpha$ and $\beta$, corresponding to the low- and high-energy spectral index. In order for the Band 
function to have a peak (and thereby a width) in the $EF_E$ representation, $\alpha$ must be greater than $-2$ while $\beta$ must be smaller than $-2$. 
To minimize effects when the values are close to their limits we apply a cut requiring $\alpha > -1.9$ and $\beta < -2.1$, with the cut in $\beta$ being
most restrictive. The reason for this is that as the peak energy approaches the upper boundary of the energy range, the high energy part of the ``turn over''
disappears leading to artificially hard or unconstrained $\beta$ values. Since our spectra are at peak flux, it is more common for the peak energy to be
at a high value and therefore far away from the low energy boundary. We further tested that our results do not depend on the exact choice of cutoff values. 
The cut reduced the BATSE sample by 359 and the GBM sample by 252 GRBs (20\% and 27\%, respectively).

From the definition in Eq.~\ref{widthdef}, it is clear that $W$ will be in ``units'' of dex. Furthermore, it is an invariant, redshift independent quantity. As the area in
the $\log EF_E$ vs $\log E$ representation indicates at what energies most of the power is radiated, $W$ will give a measure of how spread (or compact) in 
energy decades that power is.

\section{Results}

\begin{figure}
\includegraphics[width=8.4cm,angle=0]{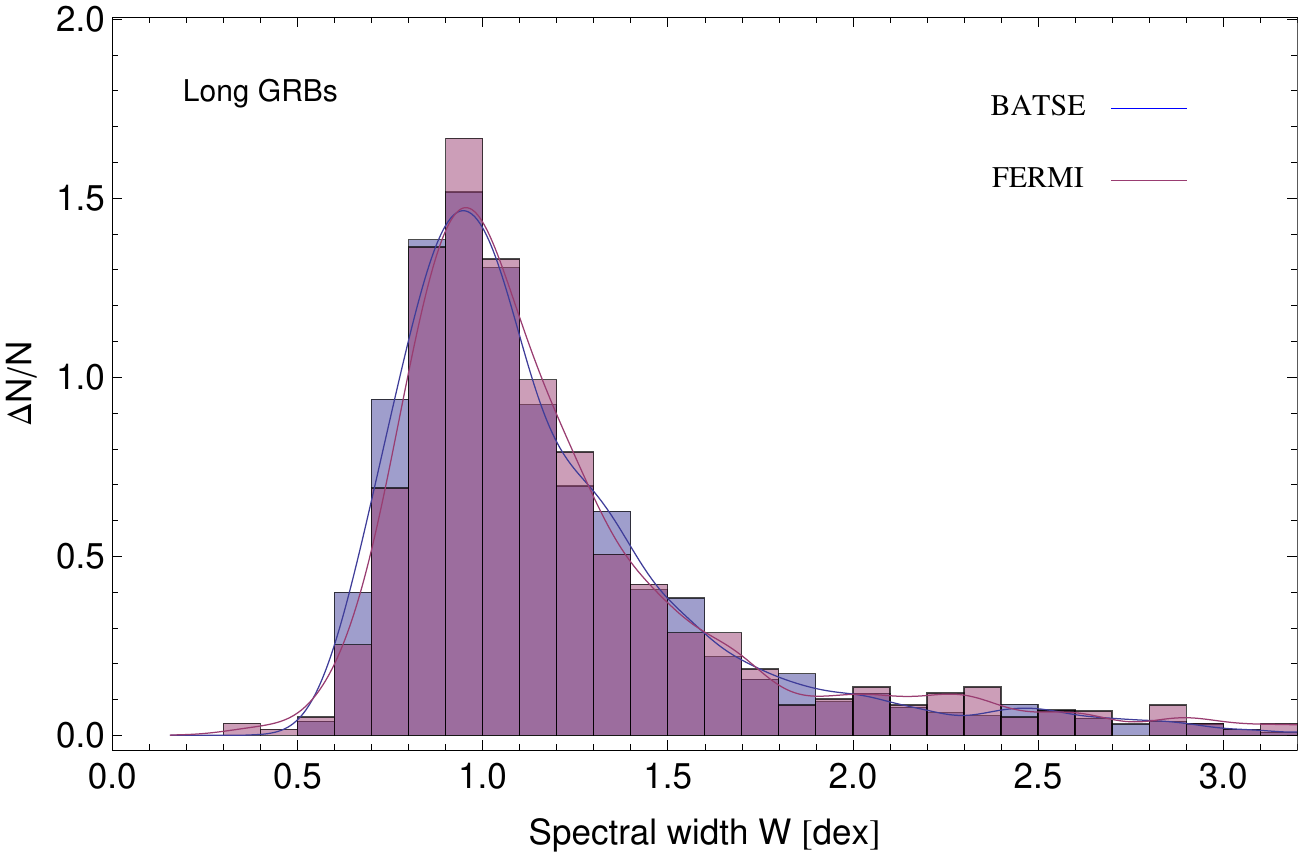}
\caption{Distribution of spectral width parameter $W$ for the sample of spectra from long GRBs. The two histograms show the 1279 BATSE bursts and the 594 
GRBs observed with {\gbm}. Solid lines show the kernel density estimation, a non-parametric way to estimate the probability density function. The distribution 
is very similar for both instruments, peaking around $W\sim1$ and a tail extending to larger widths.}
\label{longwidths}
\end{figure}

As noted above, we analyse long and short GRBs separately for each instrument. The distribution of widths for the spectra of the long GRBs are presented in 
Figure~\ref{longwidths}.

\begin{figure}
\includegraphics[width=8.4cm,angle=0]{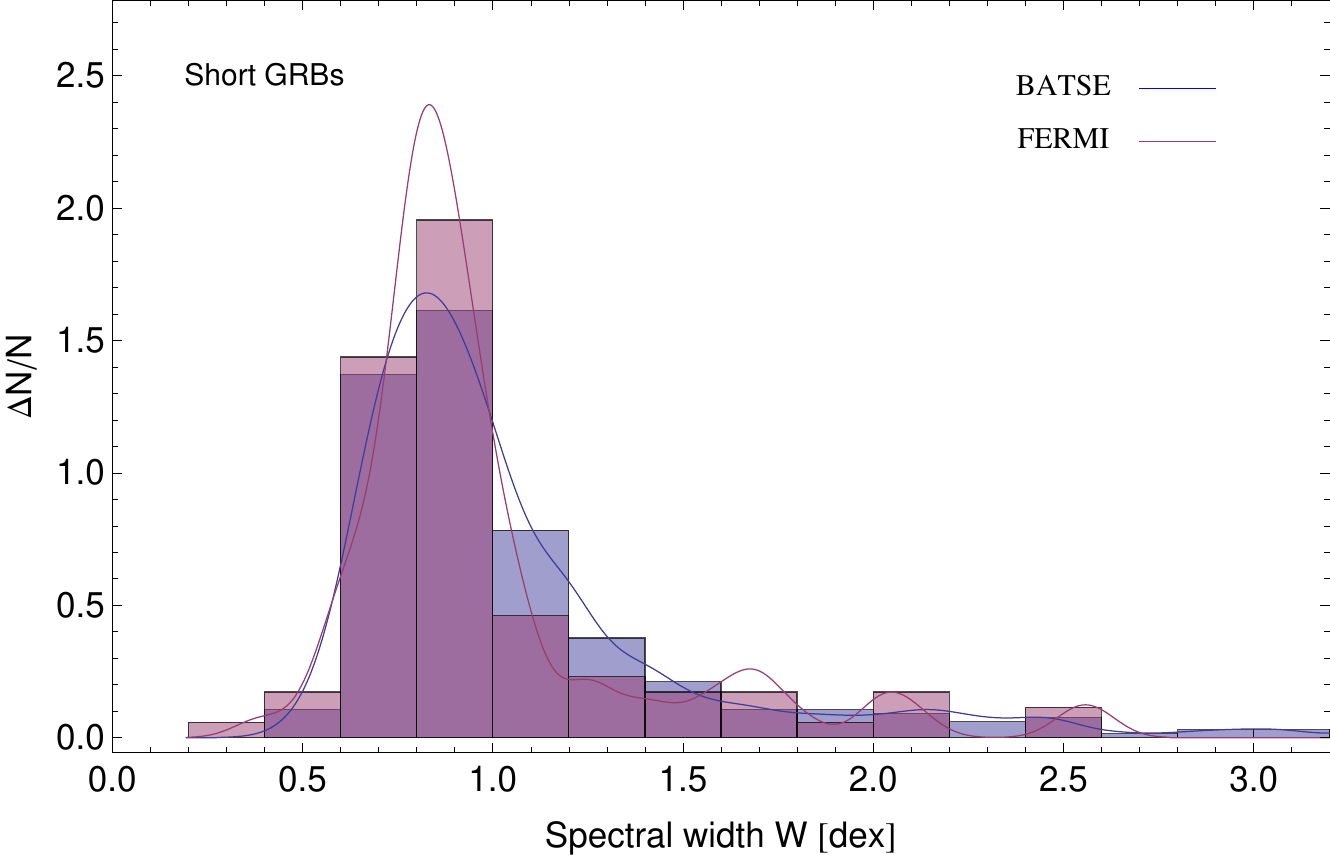}
\caption{Distribution of spectral width parameter $W$ for the sample of spectra from short GRBs. The histograms show the 332 BATSE bursts and the 87 
short GRBs observed with {\gbm}. Solid lines show the kernel density estimation.The distribution is very similar for both instruments, peaking below 
$W\sim1$ and with a tail extending to larger widths.}
\label{shortwidths}
\end{figure}

The distribution peaks at $W\sim1$, although many GRBs can be found in the tail extending towards larger widths. Virtually no spectra 
have $W<0.5$. The median value of $W$ is 1.05 for the BATSE sample and 1.07 for the GBM sample. Looking at the tail extending to larger widths, we 
find that it is for both instruments populated by spectra 
where $\beta$ is close to our cutoff value of $-2.1$. Due to this, we choose the median and quartile deviation to describe the distribution as they
are robust estimators, despite the drawback that associated probability tests have to be calculated numerically, for which we use the bootstrap 
method \citep[see e.g.,][]{press}. We note that values of $\beta>-2$ is an artificial problem arising from the limited energy window, as a peak must 
exist somewhere in the spectrum. 

Testing the GBM sample against the (larger) BATSE sample, we find that the difference between the medians is not significant. This confirms
that the distribution is not dependent on instrument, but intrinsic to the spectra themselves.

We now analyze the short GRBs. Figure~\ref{shortwidths} shows the distribution of $W$ for the short bursts in our sample, separated for the two instruments. 
As in the case of long GRBs, the distribution of $W$ shows a unimodal distribution where most of the values are clustered in a narrow range. In the case 
of the short GRBs, this peak is slightly below a value of 1. Again, the two instruments have very similar distributions, with a median of 0.91 for the BATSE 
sample and 0.86 for the GBM sample. The medians of the two distributions are not significantly different. 

The characteristics for all sample distributions are summarized in Table~\ref{locationtable}. In order to estimate the mean relative uncertainty for
individual burst widths we use Monte Carlo methods, and find that it is $\sim0.15$. 

\begin{table}
\begin{tabular}{l l l l l l}
 & \multicolumn{2}{c}{Long} & & \multicolumn{2}{c}{Short} \\
 \cline{2-3} \cline{5-6}
 & BATSE & GBM & & BATSE & GBM \\
 Sample size & 1279 & 594 & & 332 & 87\\
 Median & 1.05 & 1.07 & & 0.91 & 0.86\\
 Quartile dev. & 0.23 & 0.23 & & 0.20 & 0.12\\
\end{tabular}
\caption{Characteristics of the $W$ distributions of the two instruments, separated for long and short GRBs.}
\label{locationtable}
\end{table}

As the integration times in the GBM data are different for long and short GRBs, we compare the two classes using only the BATSE bursts. 
We find that the chance probability that the two samples come from the same distribution is less than $10^{-6}$, and therefore conclude 
that there is a highly significant difference between the two types of GRBs. If the difference were due to spectral evolution, we would expect 
short GRBs to have larger values of $W$, as most of their duration lies within the integration time (2.048\,s). The fact that short GRB spectra 
are seen to be more narrow thus strengthens the case for the difference being intrinsic. The result cannot be explained by differences in 
the peak energy between the two types of bursts, as we find no correlation between peak energy and $W$. Neither do we see any 
correlation with $T_{90}$.

In order to confirm that the difference is not due to a binning effect, we also analyze the time-integrated spectra. As expected, the
median width increases, with a larger change seen in long GRBs. These spectra show an even stronger and statistically more significant 
disparity between short and long bursts (the median is 0.96 for short GRBs and 1.26 for long).

\subsection{Comparison to emission mechanisms}

As noted above, the Band function does not provide any physical interpretation of the GRB spectrum. We have therefore calculated the width of spectra
from known physical processes. In order to make the comparison as general as possible, we select only a few ``basic" alternatives: thermal emission and
synchrotron radiation for a few electron distributions. Cooling is neglected in the synchrotron calculations, meaning the spectra are at the extreme limit for slow 
cooling. Fast cooling synchrotron spectra are significantly wider. We stress that these are not to be seen as models for GRB prompt emission; rather, their simplicity 
make them serve as fundamental limits to which any model including these processes will adhere. For each of these we have calculated $W$ according 
to Eq.~\ref{widthdef}. The results are presented in Table~\ref{width_proc}.

\begin{table}
\begin{tabular}{l l}
Process & $W$ \\
\hline
Planck function & 0.54 \\
Monoenergetic synchrotron &  0.93 \\
Synchrotron from Maxwellian e$^-$ & 1.4 \\
Synchrotron from power-law e$^-$, index $-2$ & 1.6 \\
Synchrotron from power-law e$^-$, index $-4$ & 1.4 \\
\end{tabular}
\caption{Width parameters for spectra generated by thermal emission and synchrotron emission from basic electron distributions.}
\label{width_proc}
\end{table}

\begin{figure}
\includegraphics[width=8.6cm,angle=0]{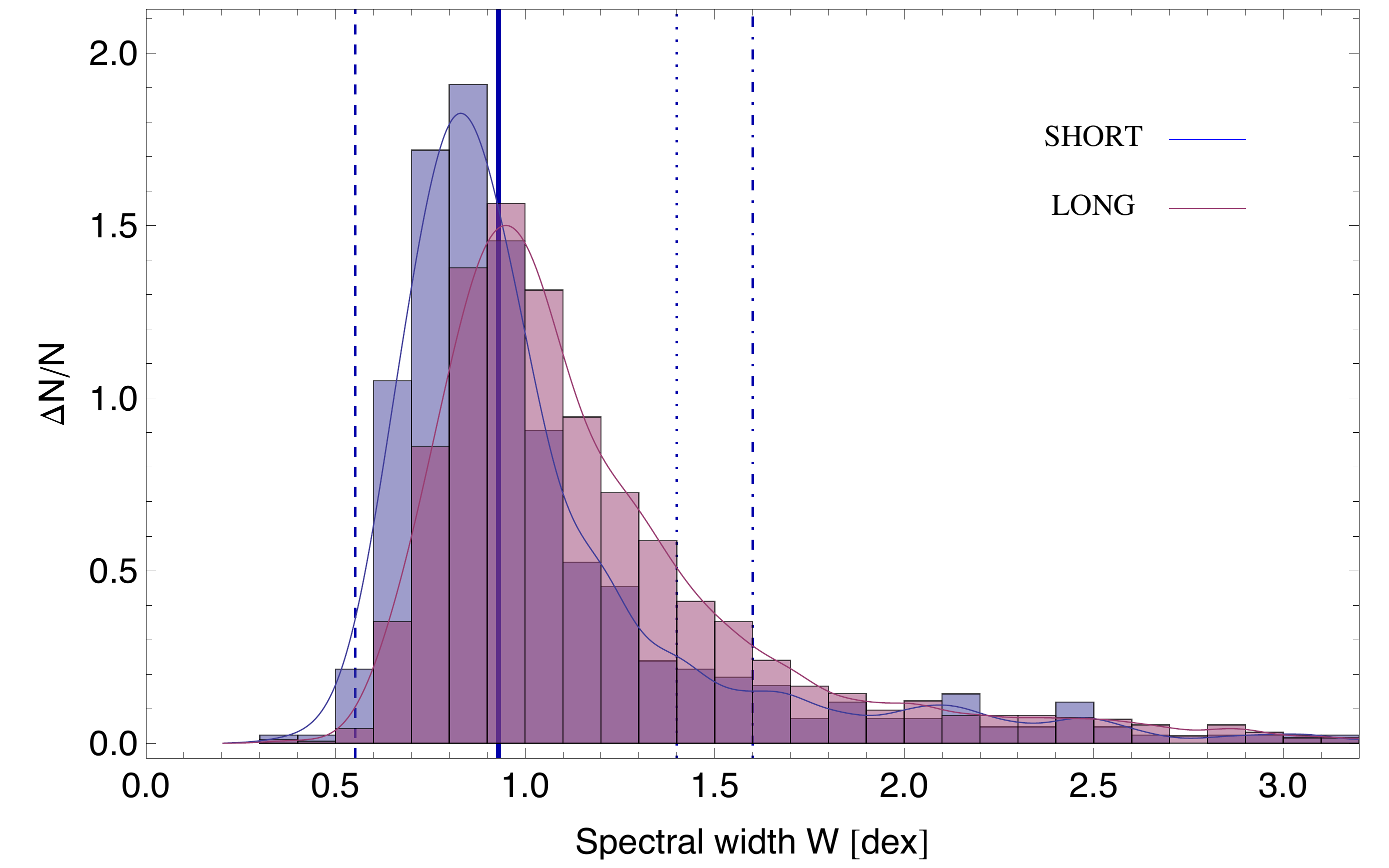}
\caption{Distribution of spectral width parameter $W$ for long and short bursts from both instruments. Thin solid lines show the kernel density estimation. The 
vertical lines indicate $W$ for spectra from the corresponding physical process: blackbody radiation (dashed), monoenergetic synchrotron (thick), synchrotron 
from Maxwellian electrons (dotted) and synchrotron from power-law electrons with index $-2$ (dash-dot).}
\label{widthcomp}
\end{figure}

To facilitate the comparison with the data, we overplot some of these limits on the data in Fig.~\ref{widthcomp}. Interestingly, the peak of the 
distribution occurs close to the width of monoenergetic synchrotron, which is not a physically realistic scenario. Figure~\ref{widthcomp} also 
shows that synchrotron emission from all electron distributions gives significantly wider spectra than generally observed. Conversely, almost 
no observed spectrum is more narrow than the Planck function (and all those are consistent with $W=0.5$ within our estimated uncertainty).

We stress that the definition in Eq.~\ref{widthdef} means that $W$ becomes a constant for the processes considered, and therefore independent 
of the location of the spectral peak. For example, a Planck function will have the same value of $W$ for all temperatures. Similarly, 
for Band function fits to the spectra, $W$ is independent of the peak energy and only depends on $\alpha$ and $\beta$. This property is particularly 
valuable, as the peak energy of GRB spectra varies throughout the burst.  

\section{Discussion}

The standard procedure of GRB analysis is to fit individual spectra with more or less physically motivated models. However, any emission mechanism
proposed must not only be able to match a single GRB spectrum, but also have the potential to reproduce the distribution in the entire observed GRB 
sample. When it comes to width of the spectrum, it is relatively easy to broaden the spectrum. For instance, spectral evolution or a combination of several
components will give a broader spectrum than predicted by a single, stable emission process. The lines in Fig.~\ref{widthcomp} may therefore be seen
as lower limits - a process can be part of any wider spectrum, but cannot be a strong component in more narrow spectra. In contrast, even though we 
have used peak flux spectra, finite integration time is needed for sufficient signal-to-noise. We can therefore not rule out that rapid spectral evolution 
has broadened the observed spectra. The measured width parameter must thus be seen as an upper limit. 

At first glance it seems striking that the peak of the width distribution occurs around the same values as the width of monoenergetic synchrotron.
However, the physical conditions required for this - constant magnetic field strength and no spread in electron energies - seem highly unreasonable.
Even if such conditions were somehow created at the onset, as the electrons radiate they would quickly cool and spread in energy. Additionally,
as the mean relative uncertainty is only 15\%, monoenergetic synchrotron could not explain the distribution down to $W\sim0.5$. We therefore
conclude that this overlap is most likely coincidental.

The fact that nearly half of the observed spectra are more narrow than monoenergetic synchrotron poses serious problems for this emission mechanism.
Assuming a distribution of electron energies makes the emitted spectrum even broader, worsening the issue. There are of course many parameters which
could in principle be varied, and such attempts have been made to alleviate the ``line of death" issue. For instance, Klein-Nishina losses can significantly 
alter the low-energy spectrum of synchrotron emission \citep{dai11}. Additionally, \citet{uz14} have shown that altering the magnetic 
field structure along the radial direction of the outflow can also modify the low-energy slope. However, the example spectra shown in \citet{uz14} all have
$W \geq 2$. In all cases, the spectrum of monoenergetic synchrotron must be seen as extreme, and any realistic assumption will give broader spectra. Based 
on this ``width of death" we therefore conclude that synchrotron emission under the standard assumption of isotropic pitch angle electron distributions cannot 
be the dominant emission process in the majority of the observed spectra. More complex models, such as small pitch angle synchrotron, can be invoked to
create more narrow spectra. However, even the most narrow spectrum in the limit of very small pitch angles has a width parameter of $W\sim1.1$ \citep{lp02}. 

It is also interesting to note that practically no spectrum is more narrow than the Planck function. This makes physical sense, as a blackbody is by definition the 
most efficient radiative process. However, there are only a few GRBs narrow enough to match the Planck function. So while thermal emission can be a component
in all observed spectra, other mechanisms must be involved. Particularly, if invoking thermal radiation a way must be found to broaden the spectrum. Such
suggestions have for instance been subphotospheric dissipation \citep{ryd11} and inverse Compton scattering \citep{peer06,bel10}. 

There have recently been several studies where GRB spectra have been fit with a combination of blackbody and Band components, with the blackbody
appearing as a subdominant peak around 100 keV \citep[e.g.,][]{gui11,axe12,bur14}. These fits tend to result in higher values of the Band function peak energy,
and softer values of $\alpha$, which may to some extent alleviate the ``line of death" problematics, and they are often interpreted as a the result of both
thermal (the blackbody) and synchrotron radiation (the Band function). However, we stress that if the spectrum comprises more than one component, one or
both of these components must be {\it more} narrow than the total spectrum. The Band component in these multi-component spectra will therefore always be 
more narrow than the spectrum as a whole, so these approaches {\it worsen} the mismatch between theoretical and observed width if assuming a synchrotron 
origin. 

The observed spread in the distributions, represented by quartile deviation in Table~\ref{locationtable}, is a combination of intrinsic spread and 
measurement uncertainty. Our estimated mean uncertainty of 0.15 thereby implies that the intrinsic spread must be quite small. This provides a strong
constraint to models explaining the emission: regardless of the emission process assumed it seems likely that there is a large range of possible parameter 
values, and one would therefore expect a correspondingly large spread in the width of the spectra.

In a physical picture, there is no reason why the dominant emission process should be the same in all GRBs. In the context of the fireball model,
it is likely that energy dissipation can occur at different places in the jet for different bursts. GRBs where the bulk of the dissipation occurs below the
photosphere will have a strong thermal component \citep[which may not need to be a blackbody; see for example][]{nym11}. In other cases the bulk of the
dissipation may occur further out in the jet giving rise to completely different spectra. We stress that although our results tend to rule out synchrotron as
the dominant process behind the majority of GRB spectra, it may still be present as a weak component. 

Our discovery that there is a significant difference between long and short GRB spectra is of particular interest, as there are very few ways that these two 
groups are known to differ. Such differences could for example be due to the central engine or differences in the jet properties.
The fact that the distributions still look relatively similar seems to indicate that the same processes are active in both types of GRBs, but that the
conditions may on average be slightly different. We will explore this further in a future study.

\section{Summary and Conclusions}

We have analysed the width of 2291 gamma-ray burst spectra from both {\it CGRO}/BATSE and {\gbm}, and find that most lie in a narrow range. 
When comparing the observations to the spectra from fundamental physical processes (thermal radiation and synchrotron emission) we find that 
synchrotron emission faces severe difficulties in explaining the data; in particular, synchrotron radiation from a distribution of electron energies 
will give a spectrum wider than the majority of observed GRB spectra. On the other hand, models of photospheric emission must include mechanisms 
to significantly broaden the emitted spectrum from a Planck function to match the data. We further find that there are significant differences between 
long and short GRBs, with the latter having more narrow spectra.

\section*{Acknowledgements}
This work was supported by the Swedish National Space Board. We thank Vah{\'e} Petrosian for valuable comments.

\end{document}